\documentclass[aps,twocolumn,preprintnumbers,amsmath,amssymb,floats,citeautoscript]{revtex4-1}
\usepackage{graphicx}
\usepackage{dcolumn}
\usepackage{bm}
\usepackage{times}
\usepackage{color}
\newcommand{\vect}[1]
{{\mbox{\boldmath $#1$}}}

\begin{document}

\title{Emergence of charge degrees of freedom under high pressure\\
in an organic dimer-Mott insulator $\beta^{\prime}$-(BEDT-TTF)$_2$ICl$_2$}

\author{K.~Hashimoto$^1$}
\author{R.~Kobayashi$^1$}
\author{H.~Okamura$^2$}
\author{H.~Taniguchi$^3$}
\author{Y.~Ikemoto$^4$}
\author{T.~Moriwaki$^4$}
\author{S.~Iguchi$^1$}
\author{M.~Naka$^5$}
\author{S.~Ishihara$^{6,7}$}
\author{T.~Sasaki$^1$}

\affiliation{$^1$Institute for Materials Research, Tohoku University, Sendai 980-8577, Japan\\
$^2$Department of Physics, Graduate School of Science, Kobe University, Kobe 657-8501, Japan\\
$^3$Saitama University, Shimo-Ohkubo 225, Saitama, 338-8570, Japan\\
$^4$SPring-8, Japan Synchrotron Radiation Research Institute, Sayo, Hyogo 679-5198, Japan\\
$^5$RIKEN Center for Emergent Matter Science (CEMS), Wako, Saitama 351-0198, Japan\\
$^6$Department of Physics, Tohoku University, Sendai 980-8578, Japan\\
$^7$CREST, JST, Chiyoda, Tokyo 102-0076, Japan\;\;\;\;\;\;\;\;\;\;}

\begin{abstract}
To elucidate the pressure evolution of the electronic structure in an antiferromagnetic dimer-Mott (DM) insulator $\beta^{\prime}$-(BEDT-TTF)$_2$ICl$_2$, which exhibits superconductivity at 14.2 K under 8 GPa, we measured the polarized infrared (IR) optical spectra under high pressure. At ambient pressure, two characteristic bands due to intra- and interdimer charge transfers have been observed in the IR spectra, supporting that this salt is a typical half-filled DM insulator at ambient pressure. With increasing pressure, however, the intradimer charge transfer excitation shifts to much lower energies, indicating that the effective electronic state changes from half-filled to 3/4-filled as a result of weakening of dimerization. This implies that the system approaches a charge-ordered state under high pressure, in which charge degrees of freedom emerge as an important factor. The present results suggest that charge fluctuation inside of dimers plays an important role in the high-temperature superconductivity.
\end{abstract}

\maketitle


\section{Introduction}
Low-dimensional layered organic conductors are a promising class of materials that exhibit high temperature superconductivity. Among them, an antiferromagnetic dimer-Mott (DM) insulator $\beta'$-(BEDT-TTF)$_2$ICl$_2$, which shows superconductivity at 14.2 K under high pressure ($\sim$8 GPa)  \cite{Taniguchi03}, has aroused great interests because its transition temperature is highest among organic superconductors known to date. At ambient pressure, $\beta'$-(BEDT-TTF)$_2$ICl$_2$ is a DM insulator with an antiferromagnetic transition temperature $T_N$ of 22 K \cite{Yoneyama99,Iguchi13}. As typically seen in bandwidth-controlled Mott transition systems such as $\kappa$-(BEDT-TTF)$_2$$X$, where $X$ = Cu$_2$(CN)$_3$, Cu[N(CN)$_2$](Br, Cl) \cite{Seo04,Dressel04,Miyagawa04}, $\beta'$-(BEDT-TTF)$_2$ICl$_2$ also exhibits superconductivity when the antiferromagnetic phase is suppressed by applying pressure. It has been, therefore, believed that antiferromagnetic spin fluctuation plays an essential role in the high-$T_c$ superconductivity \cite{Kontani03,Kino04,Nakano06}. 

The crystal structure of $\beta'$-(BEDT-TTF)$_2$ICl$_2$ consists of an alternating stack of the conductive BEDT-TTF layers and insulating ICl$_2$ layers along the $a$ axis as shown in Fig.\:1(a) \cite{Kobayashi86}. Although the charge transfer between these layers leads to a 3/4-filled band system, due to strong dimerization of two BEDT-TTF molecules, the 3/4-filled band splits into bonding and antibonding bands, resulting in an effective half-filled band system [see Figs.\:1(b) and 1(c)]. The effective model to describe the electronic state is therefore considered to be the half-filled Hubbard model, in which each dimer is regarded as a unit, and charge degrees of freedom inside of dimers are disregarded. According to band structure calculations \cite{Kobayashi86,Miyazaki03}, $\beta'$-(BEDT-TTF)$_2$ICl$_2$ has a quasi--one-dimensional (Q1D) electronic structure with a pair of slightly warped open Fermi surfaces owing to the cancelation of $t_2$ and $t_5$ [see Fig.\:1(b)]. This material, however, becomes a Mott insulator due to the strong on-dimer Coulomb repulsion $U_{d}$. Thus, $\beta'$-(BEDT-TTF)$_2$ICl$_2$ has been considered to be a typical Q1D DM insulator with one hole per each dimerized molecules at ambient pressure.

However, unlike $\kappa$-(BEDT-TTF)$_2$$X$, the phase diagram of $\beta'$-(BEDT-TTF)$_2$ICl$_2$ under pressure simply cannot be accounted for within the framework of the bandwidth-controlled Mott transition system. Recent $^{13}$C NMR measurements under pressure up to 3 GPa \cite{Eto10} have revealed that the antiferromagnetic moment of 1$\mu_B$/dimer at ambient pressure, which is a hallmark of a DM insulator, decreases to 0.5$\mu_B$/dimer at 0.6 GPa. This result implies that the DM insulator picture of $\beta'$-(BEDT-TTF)$_2$ICl$_2$ may be violated under high pressure. Indeed, first-principles band structure calculations \cite{Miyazaki03} have pointed out that the degree of dimerization is significantly reduced with increasing pressure, suggesting that the electronic structure of $\beta'$-(BEDT-TTF)$_2$ICl$_2$ changes from the effective half-filled to a 3/4-filled band system under pressure. In this situation, the effective model should be the 3/4-filled extended Hubbard model that takes into account the charge degrees of freedom inside of dimers and interdimer Coulomb interactions. Recent theoretical calculations based on the extended Hubbard model \cite{Dayal11,Sekine13} have revealed that in weakly dimerized systems both charge and spin fluctuations play an important role in unconventional superconductivity, pointing out the need to revisit the pairing mechanism that drive the high-$T_c$ superconductivity of $\beta'$-(BEDT-TTF)$_2$ICl$_2$. Since a detailed knowledge of the electronic structure is an essential starting point for understanding the superconducting pairing mechanism, experimental studies on the electronic structure of $\beta'$-(BEDT-TTF)$_2$ICl$_2$ under high pressure are highly desired. However, the high-pressure effect on the electronic structure of $\beta'$-(BEDT-TTF)$_2$ICl$_2$ have not been reported so far, except for the first-principles band structure calculations \cite{Miyazaki03}.

\begin{figure}[t]
\includegraphics[width=1.0\linewidth]{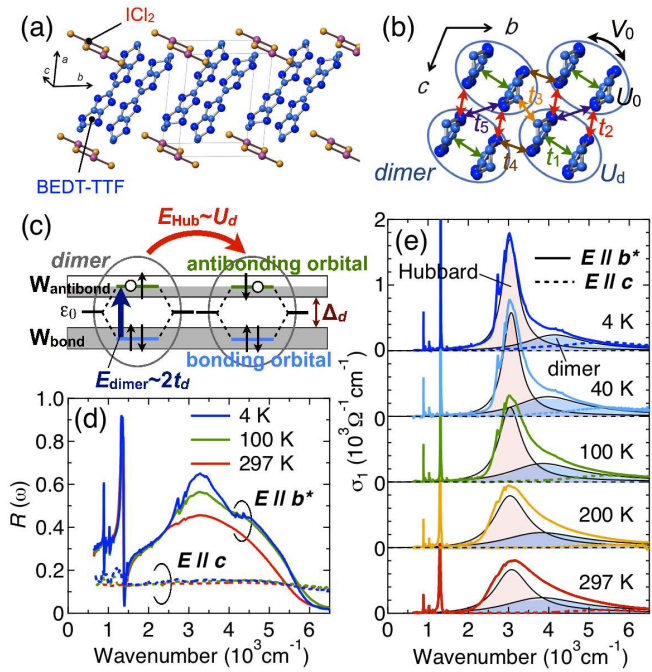}
\caption{(Color online). (a) Crystal structure of $\beta^{\prime}$-(BEDT-TTF)$_2$ICl$_2$. (b) Arrangement of BEDT-TTF molecules within the $bc$ plane viewed along the long axis of the BEDT-TTF molecule. $U_0$ is the on-molecule Coulomb interaction, $V_0$ is the intradimer Coulomb interaction, and $U_{d}$ is the effective on-dimer Coulomb interaction. The transfer integrals, $t_1\sim t_5$, are also shown. The ovals represent dimers. In the strong dimerization limit, $t_{d} \approx t_1$. (c) Schematic picture of intra- and interdimer charge transfer excitations in a DM insulator. $\Delta_d$ denotes the dimerization gap. (d) Optical reflectivity and (e) conductivity spectra of $\beta^{\prime}$-(BEDT-TTF)$_2$ICl$_2$  for ${\vect{E} \parallel \vect{b^{\ast}}}$ and ${\vect{E} \parallel \vect{c}}$ at various temperatures at ambient pressure. Solid lines in (e) indicate the Hubbard and dimer bands. The spectra in (e) are offset for clarity.
} \label{optical}
\end{figure}

The infrared (IR) optical conductivity is a powerful probe to elucidate low-energy excitations reflecting the electronic structure of matter. Especially, in strongly correlated organic compounds, characteristic electronic structures associated with the on-site and inter-site Coulomb interactions emerge in the mid-IR region \cite{Faltermeier07,Dumm09}. In this study, in order to investigate the pressure evolution of the electronic structure of $\beta'$-(BEDT-TTF)$_2$ICl$_2$, we performed polarized reflectivity measurements under high pressure by using a synchrotron radiation light source of SPring-8 and a diamond anvil cell (DAC) \cite{Matsunami09,Okamura11}. 
We show that the electronic structure of $\beta'$-(BEDT-TTF)$_2$ICl$_2$ evolves from effective half-filled to 3/4-filled with increasing pressure. Moreover, we find that the dimensionality of the electronic structure switches from Q1D to 2D, in which the high-$T_c$ superconductivity emerges.

The rest of the paper is as follows. In Sec.\:II we briefly present the experimental details of the optical conductivity experiments at ambient pressure and under high pressure. The results of the optical conductivity at ambient pressure and under high pressure are presented in Secs.\:III A and B, and discussed in the following Secs.\:III C-E. Finally, our main results are summarized in Sec.\:IV. Analytical details of the optical conductivity spectra under pressure are given in Appendix. 

\section{Experiment}
Single crystals of $\beta^{\prime}$-(BEDT-TTF)$_2$ICl$_2$ were grown by an electrochemical oxidation method \cite{Taniguchi06}. For high-precision reflectivity measurements, single crystals with very flat surfaces are required. In this study, deuterated $\beta^{\prime}$-(BEDT-TTF)$_2$ICl$_2$ single crystals, which have much clearer surfaces than the hydrogenated ones, were used. One should note that in $\beta^{\prime}$-(BEDT-TTF)$_2$ICl$_2$ substitution effects by deuterated BEDT-TTF molecules on the fundamental magnetic and optical properties are negligibly small. Polarized reflectivity measurements in the mid-IR region (600--8000 cm$^{-1}$) at ambient pressure were performed in vacuum with a Fourier transform microscope spectrometer along the $b^{\ast}$ (which is perpendicular to the $c$ axis within the $bc$ plane and corresponding to the stacking direction of the BEDT-TTF molecules) and $c$ axes. The reflectivity in the range of 8000--45 000 cm$^{-1}$ was measured at room temperature and ambient pressure. The optical conductivity was calculated by a Kramers-Kronig (KK) transformation. Above 45 000 cm$^{-1}$, the standard $\omega^{-4}$ extrapolation was used. At low frequencies, a constant extrapolation was used since the material is insulating with a large optical gap of $\sim$2000 cm$^{-1}$ (which is estimated from the dc resistivity \cite{Tajima08}). Polarized reflectivity measurements under high pressure were performed by using a diamond anvil cell (DAC) in the range of 500--8000 cm$^{-1}$. For simultaneous measurements of ${\vect{E} \parallel \vect{b^{\ast}}}$ and ${\vect{E} \parallel \vect{c}}$, two small samples with flat surfaces were placed on the culet face of the diamond anvil with a small amount of Apiezon M grease. To perform accurate measurements of the reflectivity for tiny samples, synchrotron radiation light at BL43IR in SPring-8 was used. In this study, to obtain good hydrostaticity, we used glycerin as the pressure transmitting medium, which produces more hydrostatic pressure than solid media such as KBr and NaCl \cite{Tateiwa09,Klotz12}. The pressure was monitored with the ruby fluorescence method. The reflectivity was determined by comparison with a gold thin film on the gasket. For data under pressure, Lorentz fitting analysis was used to obtain the optical conductivity spectra.

\begin{figure*}[tb]
\includegraphics[width=0.95\linewidth]{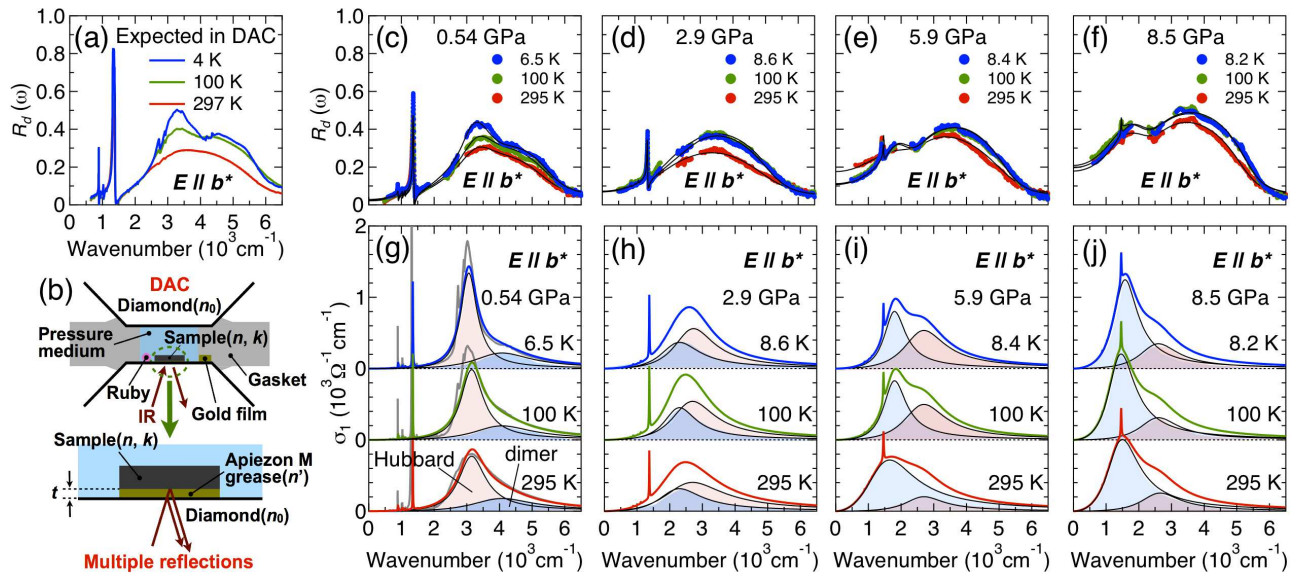}
\caption{(Color online). (a) Reflectivity spectra expected in a DAC at several temperatures for ${\vect{E} \parallel \vect{b^{\ast}}}$, which were calculated from the reflectivity spectra in Fig.\:1(d) measured in vacuum. (b) Experimental setup for the reflectivity measurements under high pressure using a DAC. (c)--(f) Reflectivity spectra measured in the DAC for ${\vect{E} \parallel \vect{b^{\ast}}}$ at 0.54 GPa, 2.9 GPa, 5.9 GPa, and 8.5 GPa, respectively. Solid lines indicate the fitting results. The spectra around 2000 cm$^{-1}$ and 3000 cm$^{-1}$ are blank owing to strong absorption by diamond and glycerin, respectively.  (g)--(j) The corresponding optical conductivity spectra for ${\vect{E} \parallel \vect{b^{\ast}}}$. The gray lines in (g) show the $\sigma_1(\omega)$ spectra measured at ambient pressure. The Hubbard and dimer bands are shown. The spectra are offset for clarity. 
} 
\end{figure*}

\section{Results and Discussion}
\subsection{Optical conductivity at ambient pressure}
The polarized reflectivity measurements in the mid-IR region at ambient pressure were performed in vacuum along the $b^{\ast}$ and $c$ axes. Figure\:1(d) shows the optical reflectivity spectra $R(\omega)$ at various temperatures for ${\vect{E} \parallel \vect{b^{\ast}}}$ and ${\vect{E} \parallel \vect{c}}$, which are consistent with previous results \cite{Kuroda88}. A large anisotropy of the reflectivity is clearly observed, reflecting the strong 1D electronic structure of $\beta^{\prime}$-(BEDT-TTF)$_2$ICl$_2$. Several vibrational features can be seen below $\sim$1500 cm$^{-1}$; these are attributed to the totally symmetric $a_g$ vibrational modes of the BEDT-TTF molecule. In the mid-IR region, $R(\omega)$ for ${\vect{E} \parallel \vect{b^{\ast}}}$ shows broad features, as typically seen in a DM insulator. The corresponding optical conductivity spectra $\sigma_1(\omega)$ obtained from the KK transformation are plotted in Fig.\:1(e). At room temperature, $\sigma_1(\omega)$ for ${\vect{E} \parallel \vect{b^{\ast}}}$ has a broad and asymmetric shape in the range of 2000--6000 cm$^{-1}$. As lowering the temperature, however, a peak structure centered at 3000 cm$^{-1}$ becomes sharper, which makes it clearer that the spectra for ${\vect{E} \parallel \vect{b^{\ast}}}$ consists of two bands located at 3000 cm$^{-1}$ and 4000 cm$^{-1}$. In sharp contrast, $\sigma_1(\omega)$ for ${\vect{E} \parallel \vect{c}}$ exhibits very low conductivity, reflecting the 1D nature of the electronic structure as was expected in the reflectivity spectra.

In a DM insulator, there are two absorption bands in the mid-IR region: the so-called Hubbard and dimer bands corresponding to the inter- and intradimer charge transfers, respectively [see Fig.\:1(c)]. The excitation energy of the interdimer charge transfer, $E_{\rm{Hub}}$, is given by the effective on-dimer Coulomb interaction, $U_{d}$, which causes splitting of the antibonding band.
In the extended Hubbard model, the effective on-dimer Coulomb interaction, $U_{d}$, is represented by  $U_{d} = 2 t_{d} +(U_0+V_0)/2-\sqrt{4t_{d}^2+(U_0-V_0)^2/4}$, where $U_0$ is the on-molecule Coulomb interaction, $V_0$ is the intradimer Coulomb interaction, and $t_{d}$ is the intradimer transfer integral \cite{Fortunelli97}. On the other hand, the excitation energy of the intradimer charge transfer, $E_{\rm{dimer}}$, is given by $2 t_{d}$ in the strong dimerization limit ($U_0$, $V_0$ $\ll$ $t_d$). Recent band structure calculations based on the maximally localized Wannier orbitals (MLWO) for a series of dimerized hall-filled organic salts \cite{Koretsune14} have a remarkable development in estimates of model parameters. The calculations based on the MLWO for $\beta^{\prime}$-(BEDT-TTF)$_2$ICl$_2$ have estimated the intradimer charge transfer $2 t_{d}$ $\approx$ $2 t_1$ to be 500 meV (which corresponds to approximately 4000 cm$^{-1}$), showing an excellent agreement with the experimental value of 4000 cm$^{-1}$. Therefore the band at 4000 cm$^{-1}$ is assigned to the dimer band. On the other hand, the band at 3000 cm$^{-1}$ is characterized by the sharpness; it's much sharper than that of the 2D $\kappa$-type system \cite{Faltermeier07,Dumm09}, which reflects the narrow density of states in the 1D Mott insulator \cite{Tajima08}. This fact strongly supports that the band at 4000 cm$^{-1}$ is assigned to the Hubbard band. In the strong dimerization limit, the central frequency of the Hubbard band becomes smaller than that of the dimer band [see Fig.\:4(e) for details]. Therefore, the present result that the excitation energy of the Hubbard band transition is lower than that of the dimer band transition indicates that $\beta^{\prime}$-(BEDT-TTF)$_2$ICl$_2$ is located in the strong dimerization regime at ambient pressure.

\subsection{Optical conductivity under pressure}
In order to elucidate the pressure evolution of $\sigma_1(\omega)$, the polarized reflectivity measurements under high pressure were performed by using a DAC. In a DAC, the reflectivity is measured at a sample/diamond interface, not at a sample/vacuum interface [see Fig.\:2(b)]. Therefore the large refractive index of diamond ($\sim$2.4) makes a difference between the spectra measured in a DAC and vacuum, $R_d(\omega)$ and $R(\omega)$ (see Appendix for details). In addition to this, multiple reflections in a thin layer of Apiezon M grease between the sample and diamond also affects the $R_d(\omega)$ spectra. Our analysis considering the above corrections have shown that the thickness of the thin layer of Apiezon M grease is about 300-400 nm. Figure\:2(a) shows the $R_d(\omega)$ spectra for ${\vect{E} \parallel \vect{b^{\ast}}}$ expected from $R(\omega)$ measured in vacuum at ambient pressure, which well reproduces $R_d(\omega)$ actually measured at a low pressure of 0.54 GPa [see Fig.\:2(c)].

\begin{figure}[t]
\includegraphics[width=1.0\linewidth]{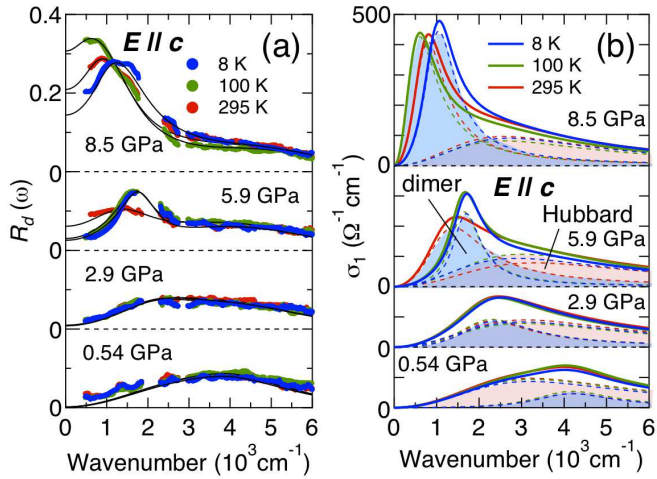}
\caption{(Color online). (a) Reflectivity spectra measured at various pressures for ${\vect{E} \parallel \vect{c}}$. Solid lines indicate the fitting results. (b) Optical conductivity spectra obtained from the fitting analysis for ${\vect{E} \parallel \vect{c}}$. The spectra are offset for clarity.
} \label{sigma_irradiation}
\end{figure}

To extract $\sigma_1(\omega)$ from the measured $R_d(\omega)$ spectra, we fit $R_d(\omega)$ to a Lorentz oscillator model (see Appendix for details). We show examples of the fitting for ${\vect{E} \parallel \vect{b^{\ast}}}$ and $\vect{E} \parallel \vect{c}$ in Figs.\:2(c)-(f) and Fig.\:3(a), respectively. The measured $R_d(\omega)$ is well reproduced by two Lorentz oscillators in the mid-IR region, as well as several sharp phonons below $\sim$1500 cm$^{-1}$ in the case of ${\vect{E} \parallel \vect{b^{\ast}}}$. The corresponding optical conductivity spectra for ${\vect{E} \parallel \vect{b^{\ast}}}$ and $\vect{E} \parallel \vect{c}$ obtained from the fitting analysis are shown in Figs.\:2(g)-(j) and Fig.\:3(b), respectively. At 0.54 GPa, $\sigma_1(\omega)$ for ${\vect{E} \parallel \vect{b^{\ast}}}$ exhibits very similar behavior to that of ambient pressure; the Hubbard and dimer bands have been observed in the mid-IR region. With further increase in pressure, the Hubbard band becomes broader as shown in Figs.\:2(h)-(j), but the central frequency doesn't change so much with pressure. In sharp contrast, the dimer band shifts to much lower frequencies between 0.54 GPa and 2.9 GPa, and with increasing pressure the spectral weight (SW) becomes much larger.

\begin{figure*}[tb]
\includegraphics[width=0.6\linewidth]{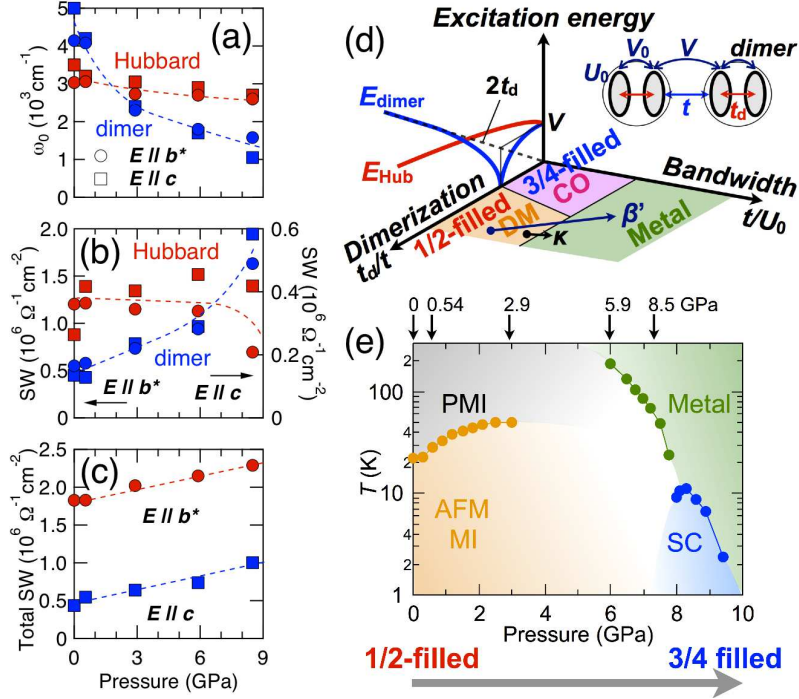}
\caption{(Color online). (a)--(c) Pressure dependences of the central frequencies, SWs, and total SWs of the Hubbard and dimer bands for ${\vect{E} \parallel \vect{b^{\ast}}}$ and $\vect{E} \parallel \vect{c}$ at the lowest temperature, respectively. (d) The excitation energies of the Hubbard and dimer bands in the strong Coulomb interaction limit are shown as a function of degree of dimerization. The schematic phase diagram expected in the extended Hubbard model with the molecular dimer degree of freedom and the effective model Hamiltonian derived from it is also shown in the bottom plane as functions of degree of dimerization and bandwidth, where the Coulomb interaction parameters are fixed. Here, the on-molecular Coulomb interaction $U_0$, the intra-dimer Coulomb interaction $V_0$, the inter-dimer Coulomb interaction $V$, the intra-dimer transfer integral $t_d$, and the inter-dimer transfer integral $t$ are taken into account. The Hartree-Fock type approximation is adopted in the analyses of the extended Hubbard model. ``Dimerization" and ``Bandwidth" axes in the phase diagram correspond to the parameters $t_d/t$ and $t/U_0$, respectively. (e) The $p$-$T$ phase diagram of $\beta^{\prime}$-(BEDT-TTF)$_2$ICl$_2$. $T_{\rm{MI}}$ (green circle) and $T_c$ (blue circle) were determined by the resistivity measurements \cite{Taniguchi03}. $T_N$ (orange circle) was obtained from the $^{13}$C NMR measurements under pressure up to 3 GPa \cite{Eto10}. The magnetic properties above 3 GPa (the blank area) have yet to be determined. The arrows depict pressure values studied here.
} 
\end{figure*}

\subsection{Pressure-induced modulation of the degree of dimerization}
To clarify the pressure evolutions of the Hubbard and dimer bands, we plot the pressure dependence of the central frequencies and SWs of each band in Figs.\:4(a) and 4(b). Between 0.54 GPa and 2.9 GPa, the central frequency of the dimer band rapidly decreases. In contrast, the pressure dependence of the Hubbard band is gentle compared to that of the dimer band. As a result, above 2.9 GPa, the central frequency of the dimer band becomes lower than that of the Hubbard band. As mentioned above, in the strong dimerization limit, the central frequency of the Hubbard band should be smaller than that of the dimer band. The present result therefore indicates that the simple DM insulator picture of $\beta^{\prime}$-(BEDT-TTF)$_2$ICl$_2$ has been violated above 2.9 GPa. 

Recent theoretical calculations based on the extended Hubbard model \cite{Naka10,Naka13} have suggested that by weakening the degree of dimerization or by increasing the interdimer Coulomb interaction, the DM insulating state can be suppressed and evolve into a CO insulating state, where the electronic charge distributions are polarized inside of dimers [see the schematic phase diagram of Fig.\:4(d)]. According to the calculations, the intradimer charge transfer corresponding to the dimer band transition significantly softens compared to $2t_{d}$ approaching the phase boundary between the DM and CO insulating phases, resulting in a collective excitation \cite{Itoh13, explanation}. On the other hand, the Hubbard band transition smoothly evolves into a charge transfer excitation induced by the inter-site Coulomb interaction $V$, as indeed observed in quarter-filled organic conductors in a CO phase \cite{Dressel03,Drichko06,Kaiser10,Hashimoto14}. Very importantly, in this situation, the excitation energy of the dimer band transition can be smaller than that of the Hubbard band transition. In the light of this fact, the present result for $\beta^{\prime}$-(BEDT-TTF)$_2$ICl$_2$ that the excitation energy of the dimer band transition becomes smaller than that of the Hubbard band under pressure strongly suggests that the electronic structure of $\beta^{\prime}$-(BEDT-TTF)$_2$ICl$_2$ changes from effective half-filled to 3/4-filled as a result of weakening of dimerization with increasing pressure. Such a decrease of dimerization under pressure has been discussed also in (TMTTF)$_2$$X$ salts \cite{Rose13,Jacko13}.

Under pressure, in addition to weakening of dimerization, an increase of the bandwidth has been observed. Figure\:4(c) shows the pressure dependence of the total SWs for ${\vect{E} \parallel \vect{b^{\ast}}}$ and $\vect{E} \parallel \vect{c}$ at the lowest temperature. The total SW continuously increases with increasing pressure for both polarization directions, reflecting the increase in the bandwidth due to the increase in the transfer integrals. Therefore in $\beta^{\prime}$-(BEDT-TTF)$_2$ICl$_2$, when the pressure is applied, the system would follow the track as sketched in Fig.\:4(d). What is remarkable here is that the system crosses the vicinity of the CO phase, which can be associated with the superconducting pairing mechanism of $\beta^{\prime}$-(BEDT-TTF)$_2$ICl$_2$ as discussed later. In the case of $\kappa$-(BEDT-TTF)$_2$$X$, a slight application of pressure ($<$ 0.1 GPa) induces a Mott transition. Thus, the degree of dimerization is almost unchanged in the low-pressure region, so that the system remains in the strong dimerization limit. Indeed, this has been confirmed by the chemical pressure studies for $\kappa$-(BEDT-TTF)$_2$Cu[N(CN)$_2$]Br$_x$Cl$_{1-x}$ \cite{Faltermeier07,Dumm09}, in which the central frequencies of neither the Hubbard nor the dimer bands change with $x$, indicating that the system is located far away from the CO phase. 

\subsection{Dimensional crossover of the electronic structure under pressure}
Another important finding in this study is that the SW for $\vect{E} \parallel \vect{c}$ significantly increases with increasing pressure. The total SW for $\vect{E} \parallel \vect{c}$ at 8.5 GPa becomes twice compared to that at ambient pressure, whereas for ${\vect{E} \parallel \vect{b^{\ast}}}$ it increases only by 25\%. Consequently, the anisotropy of the total SW becomes smaller. This result indicates that the dimensionality of the electronic structure of $\beta^{\prime}$-(BEDT-TTF)$_2$ICl$_2$ evolves from Q1D to 2D with increasing pressure. 
According to the previous theoretical studies based on the fluctuation-exchange (FLEX) approximation including only the on-site Coulomb interaction \cite{Kontani03,Kino04,Nakano06}, with increasing pressure the nesting condition of the Fermi surface becomes degraded owing to the dimensional crossover of the Fermi surface. As a result, the magnetic structure described by the localized spins in the strongly correlated Mott insulating state changes into an itinerant SDW state. With further increase in pressure, $T_N$ decreases and the antiferromagnetic phase can be suppressed. Finally, unconventional superconductivity with $d_{xy}$-wave symmetry emerges. Thus, the origin of the superconductivity of $\beta^{\prime}$-(BEDT-TTF)$_2$ICl$_2$ is considered to be associated with antiferromagnetic spin fluctuation. However, in the above calculations, intra- and interdimer Coulomb interactions are not included. Our present study highlights the 3/4-filled nature of $\beta^{\prime}$-(BEDT-TTF)$_2$ICl$_2$ under pressure, in which the intradimer charge degrees of freedom and interdimer Coulomb interactions become more important. Recent theoretical calculations analyzed by the random-phase approximation (RPA) and FLEX based on the extended Hubbard model \cite{Sekine13} have pointed out that the superconducting phase appears near the charge-density-wave (CDW) phase as well as the SDW phase. While the extended $s$-wave pairing symmetry is favored near the CDW phase, $d_{xy}$-wave superconductivity is realized due to spin fluctuation near the SDW phase. Since both superconducting gap symmetries are classified by $A_{1g}$ irreducible representation, the two superconducting states are cooperative with each other. Then charge fluctuation inside of dimers can enhance the spin-fluctuation induced superconductivity. In this context, the present experimental results for $\beta'$-(BEDT-TTF)$_2$ICl$_2$ suggest that in addition to antiferromagnetic spin fluctuation, charge fluctuation plays an important role in the high-$T_c$ superconductivity of $\beta'$-(BEDT-TTF)$_2$ICl$_2$. 

\subsection{$p$-$T$ phase diagram of $\beta'$-(BEDT-TTF)$_2$ICl$_2$}
Finally, we compare the $p$-$T$ phase diagrams obtained in the present study and the previous resistivity measurements using a cubic anvil cell \cite{Taniguchi03}. According to the resistivity measurements, $\beta^{\prime}$-(BEDT-TTF)$_2$ICl$_2$ shows superconductivity above $\sim$7 GPa, and $T_c$ reaches 14.2 K at 8.2 GPa [see Fig.\:4(e)]. In the present study, we have focused on the mid-IR region, so we cannot discuss details of the electronic structure near the Fermi level such as a Drude peak. However, the optical response in the low-energy region below $\sim$1000 cm$^{-1}$ seems to capture the features associated with the electronic structure near the Fermi level. Below 2.9 GPa, $R_d(\omega)$ in the low-energy region doesn't change so much with temperature, reflecting the large optical gap independent of temperature. At 5.9 GPa, however, $R_d(\omega)$ in the low-energy region slightly increases with decreasing temperature from room temperature, and below 200 K it decreases with decreasing temperature and becomes temperature-independent below 100 K. This result indicates that the metal-insulator (MI) transition at 5.9 GPa occurs at about 200 K, which is consistent with the $p$-$T$ phase diagram obtained from the resistivity measurements. At 8.5 GPa, $R_d(\omega)$ in the low-energy region increases with decreasing temperature and exhibits a maximum at 80 K, which suggests that the MI transition occurs at $\sim$80 K. This implies that the highest pressure we reached corresponds to about 7 GPa in the $p$-$T$ phase diagram obtained from the resistivity measurements. The discrepancy between the present study and the resistivity measurements may come from a difference in the hydrostatic stress conditions.

The magnetic properties of the low-temperature magnetically ordered phase have been investigated up to 3 GPa by the $^{13}$C NMR measurements \cite{Eto10}. Although we measured the optical conductivity down to low temperatures at each pressure, a clear difference in the optical spectra above and below the magnetic transition has not been observed. This is because the electronic excitations associated with the Hubbard and dimer bands are not largely affected by the magnetic order. The experimental result that there is no large temperature variation at low temperatures in the optical spectra at 2.9 GPa and 5.9 GPa suggests that the system doesn't enter an insulating state such as a CO phase that drastically changes the electronic structure in the mid-IR region. Therefore the antiferromagnetic Mott insulating phase or the SDW phase should be realized in the low-temperature region above 3 GPa, as discussed above. Further experimental studies on the magnetic properties under higher pressures are desired.

\section{Conclusions}
In summary, we have performed optical conductivity measurements for $\beta^{\prime}$-(BEDT-TTF)$_2$ICl$_2$ under high pressure. At ambient pressure, two characteristic bands due to intra- and interdimer charge transfers have been observed in the optical spectra only for the polarization along the stacks of BEDT-TTF molecules, strongly supporting that this salt is a Q1D DM insulator at ambient pressure. With increasing pressure, however, the intradimer charge excitation shifts to much lower frequencies as a result of weakening of dimerization, implying that the electronic structure evolves from effective half-filled into 3/4-filled under pressure. Moreover, we have shown that the dimensionality of the electronic structure switches from Q1D to 2D, in which the high temperature superconductivity emerges. These results indicate that the phase diagram of $\beta^{\prime}$-(BEDT-TTF)$_2$ICl$_2$ is different from that of $\kappa$-(BEDT-TTF)$_2$$X$ where the superconducting phase appears in the effective half-filled band system. The present results suggest that the system approaches the charge-ordered state under pressure, in which the possible charge-fluctuation-mediated high-$T_c$ superconductivity emerges.

\section*{Acknowledgments}
We thank C. Hotta, S. Iwai, and A. Kawamoto for fruitful discussions. Synchrotron radiation measurements were performed at SPring-8 with the approvals of JASRI (2013A0089, 2013A1118, 2013B0089, and 2013B1144). This work is supported by a Grant-in-Aid for Scientific Research (Grants No. 23540409, No. 24540357, No. 25287080, No. 26287070, No. 26800173, and No. 15H00984) from MEXT and JSPS, Japan.

\section*{Appendix: Analysis of optical conductivity spectra in a diamond anvil cell}

In a DAC, the reflectivity is measured at a sample/diamond interface. According to the Fresnel's formula, the reflectivity spectra $R(\omega)$ at an interference between a sample and a transparent medium is given by 
\begin{align}
R(\omega) = \frac{[n(\omega)-n_0]^2+k(\omega)^2}{[n(\omega)+n_0]^2+k(\omega)^2}, \tag{A1}
\end{align}
where $n(\omega)$ and $k(\omega)$ are the real and imaginary parts of the complex refractive index of the sample, respectively, and the refractive indices $n_0$ for vacuum and diamond are 1 and 2.419, respectively. Therefore $R_d(\omega)$ measured in a DAC generally becomes smaller than $R(\omega)$ measured in vacuum. In addition to this, multiple reflections in a thin layer of Apiezon M grease between the sample and diamond [see Fig.\:2(b) in the main text] also affects $R_d(\omega)$, especially in the high-energy region \cite{Matsunami09,Okamura11}; this is because shorter wavelength lights are more likely to be influenced by such multiple reflections. The reflectivity considering the multiple reflections can be given by \cite{Heavens91}
\begin{align}
R_d = \left| \frac{(n_0-n^{\prime})(n^{\prime}+\hat{n})e^{i\delta}+(n_0+n^{\prime})(n^{\prime}-\hat{n})e^{-i\delta}}{(n_0+n^{\prime})(n^{\prime}+\hat{n})e^{i\delta}+(n_0-n^{\prime})(n^{\prime}-\hat{n})e^{-i\delta}}\right|^2, \tag{A2}
\end{align}
where $\delta = 2\pi\omega n^{\prime}t$, $\hat{n} = n + i k$ is the complex refractive index of the sample, $n_0=2.42$ and $n^{\prime}=1.77$ are the refractive indices of diamond and Apiezon M grease, and $t$ is the thickness of Apiezon M grease. Our analysis considering the above equation have shown that the thickness of the thin layer is about 300 nm and 430 nm for the ${\vect{E} \parallel \vect{b^{\ast}}}$ and ${\vect{E} \parallel \vect{c}}$ samples, respectively. Figure\:2(a) in the main text shows the $R_d(\omega)$ spectra for ${\vect{E} \parallel \vect{b^{\ast}}}$ expected from $R(\omega)$ measured in vacuum at ambient pressure, which well reproduces $R_d(\omega)$ actually measured at a low pressure of 0.54 GPa [see Fig.\:2(c)]. Here, to calculate $R_d(\omega)$ expected in a DAC at ambient pressure, $n(\omega)$ and $k(\omega)$ were derived from the KK analysis of $R(\omega)$ measured in vacuum at ambient pressure. Then, the obtained parameters were substituted into Eq.\;(A2) with $t=300$ nm.

To extract $\sigma_1(\omega)$ from the measured $R_d(\omega)$ spectra, we fit $R_d(\omega)$ to a Lorentz oscillator model. In this model, the real and imaginary parts of the complex dielectric function,  $\epsilon_1(\omega)$ and $\epsilon_2(\omega)$, are given by
\begin{align} 
\epsilon_1(\omega) = \epsilon_{\infty}+\sum_{i} \omega_{p,i}^2\frac{(\omega_{0,i}^2-\omega^2)}{(\omega_{0,i}^2-\omega^2)^2+\gamma_i^2\omega^2}, \tag{A3}\\
\epsilon_2(\omega) = \sum_{i} \omega_{p,i}^2\frac{\gamma_i\omega}{(\omega_{0,i}^2-\omega^2)^2+\gamma_i^2\omega^2}, \tag{A4}
\end{align}
where $\epsilon_{\infty}$ is the high-frequency dielectric constant, $\omega_{p,i}$, $\omega_{0,i}$, and $\gamma_i$ are the plasma frequency, the central frequency, and the linewidth of the $i$-th Lorentz oscillator, respectively. The real and imaginary parts of the complex refractive index of the sample, $n(\omega)$ and $k(\omega)$, are described by the above Lorentzian parameters through the relations $\epsilon_1(\omega)=n^2(\omega) - k^2(\omega)$ and $\epsilon_2(\omega) = 2n(\omega)k(\omega)$. Then, by fitting the measured $R_d(\omega)$ based on Eq.\;(A2), we can obtain the Lorentzian parameters. Thus, we can extract $\sigma_1(\omega)$ through the relation $\sigma_1(\omega) = \omega \epsilon_2(\omega)/(4\pi)$. We show the fitting results for ${\vect{E} \parallel \vect{b^{\ast}}}$ and $\vect{E} \parallel \vect{c}$ in Figs.\:2(c)-(f) and Fig.\:3(a), respectively. The corresponding optical conductivity spectra for ${\vect{E} \parallel \vect{b^{\ast}}}$ and $\vect{E} \parallel \vect{c
 }$ are shown in Figs.\:2(g)-(j) and Fig.\:3(b), respectively.



\begin{thebibliography}{99}
\bibitem{explanation}
In $\beta^{\prime}$-(BEDT-TTF)$_2$ICl$_2$, there exists only one equivalent dimer in the unit cell. Therefore only the acoustic mode can exist, in which polarizations in two adjacent dimers align in parallel with each other. Since the direction of polarization is expected to be evolved along the direction from the center of one BEDT-TTF molecule to the other one in the dimer (which is close to the $b^{\ast}$ axis), the acoustic mode appear nearly along the $b^{\ast}$ axis. However, the direction of polarization and the $c$ axis are not perfectly perpendicular to each other, so the same acoustic mode should be observable in the polarization of $\vect{E} \parallel \vect{c}$ in $\beta^{\prime}$-(BEDT-TTF)$_2$ICl$_2$.
\end{thebibliography}
\end{document}